\documentclass[11pt]{article}
\usepackage{graphicx}
\begin {document}
\begin{center}
\LARGE
High-brightness ultra-cold metastable neon-beam
\vspace{20pt}

\normalsize
Fujio Shimizu
\vspace{11pt}

Institute for Laser Science, University of Electro-Communications, 1-5-1 Chofugaoka, Chofu, Tokyo 113-8585, Japan
\vspace{11pt}

{\bf ABSTRACT}
\vspace{11pt}
\end{center}

This paper presents detailed characteristics of an ultra-cold bright meta-stable
 neon atomic beam which we have been using  for atom-interferometric 
applications.
The basis of the device is an atomic beam released from a magneto-optical
trap (MOT) which is operated with a high intensity trapping laser, high 
magnetic quadrupole field, and  large laser detuining.
An idealized MOT is that atoms stay in a high magnetic-field-seeking state
and are reflected back towards inner side by selectively absorbing the
laser photon which travels outwards of the trap. In three dimensional 
configuration, however, an atom has always finite probability to be pumped into
a low-field-seeking state,  because the direction of the magnetic field and
the propagation direction of the laser cannot be parallel in entire space. 
Furthermore, at the center of the quadrupole field there is no preferred
polarization for absorption. Mainly due to those effects a bright 
small spot of atoms is formed near the center under an appropriate operating
condition. We obtained the minimum trap diameter of 
50~$\mu$m, the atomic density nearly
$10^{13}$~cm$^{-3}$, and the atomic temperature slightly less than the
Doppler limited temperature of 200~$\mu$K. By releasing trapped atoms we obtained an bright cold atomic
beam which is not far from the collision limited atomic density.

\section {Introduction}


\begin{figure}[htbp]
\begin{center}
\includegraphics[width=10cm]{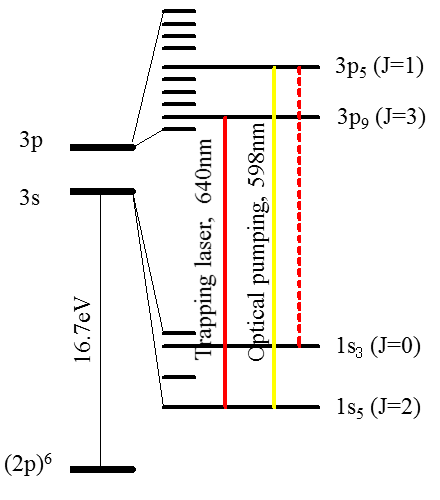}
\caption{Energy levels of a neon atom relevant to the atomic beam generation.
Neon atoms in the 1s$_5$ level are laser cooled and trapped using the cyclic
transition of 1s$_5$~-~3p$_9$ at 640~nm. The trapped atoms are illuminated
by a laser which is 
resonant to the 1s$_5$~-~3p$_5$ transition. Approximately a half 
of the 3p$_5$ atoms decay to the second metastable state 1s$_3$. The remaining 
half immediately decay to the ground state via two $J=1$ states 
by emitting a 70~nm VUV photon.
The lifetime of the 1s$_5$ and 1s$_3$ are approximately 20~s. 
}
\label{levels}
\end{center}
\end{figure}


Interferometric manipulation of atoms is still in a primitive stage 
compared to that of photons. 
When we try to work with atoms, the most serious problem is that a 
large inter-atomic 
interaction limits the atomic density 
in the beam by
many orders of magnitude less than that of photons in an optical beam.
The brightness of a classical optical  beam is proportional to
the inverse of the source area, if the total intensity is the same.
The invention of lasers removed this restriction, because in principle 
the wave front
of a laser light can be controlled at will.
Bose-Einstein condensate(BEC) of atoms\cite{cornell,ketterle} 
can provide the same benefits for an atomic beam.   
However, up to now, an atomic beam generated from BEC is operated only
intermittently. In addition, BEC is not a collection of free atoms
such as photons of a laser. It is uncertain if the atoms can be expanded 
to a coherent wave of independent atoms whose wave front can be
controlled as intended.
Therefore, it seems to be more practical to use the same technique which
has been used in classical optics. This is to emit the atomic 
beam from a small
bright source with a small energy dispersion.

The apparatus we describe below uses a "magnetic trap" formed
in the magneto-optical trap (MOT)\cite{mot}, which, we found, 
is generated stably 
when MOT is operated with high laser intensity, large detuning, and high
magnetic gradient.  The atomic species we use is Ne$^{20}$ 
in the 1s$_3$ metastable state which has no magnetic moment $F=0$. 
(Fig.~\ref{levels}) 
First, neon atoms in the 1s$_5$ metastable state ($J-2$) 
is trapped using a four beam
magneto-optical trap\cite{tetra}. Then, atoms are released continuously from the trap using
an optical pumping laser at 597,7 nm that transfers 1s$_5$ atoms to the 1s$_3$ 
state via 3p$_5 (J-1)$ state. 
The same technique can be used for other rare gas atoms such as Ar, Kr and Xe.

\begin{figure}[htbp]
\begin{center}
\includegraphics[width=8cm]{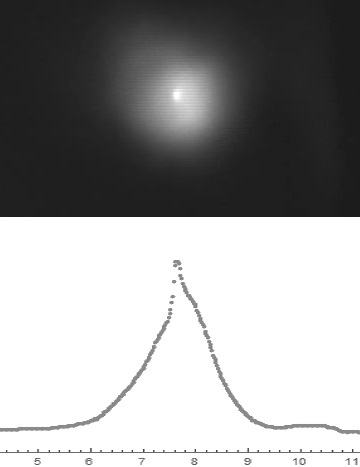}
\caption{Typical trap pattern (top) and its intensity distribution (bottom)
when the feeding of atoms into the MOT 
is sufficiently strong. When the feeding is weak 
only the central bright spot is observed. 
In this figure the  brightness of the central spot is saturated. 
Numbers below the curve is in mm. Neon atoms arrive from left.
}
\label{feeding}
\end{center}
\end{figure}

\begin{figure}[htbp]
\begin{center}
\includegraphics[width=8cm]{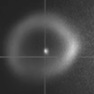}
\caption{"Magnetic trap" appears not only when the trap pattern is
symmetric. When the laser detuning is large, a slight mis-alignment
tends to form a ring shape trap. Even in this situation a bright spot 
around the $B=0$ point appears frequently. Its position is
 independent of the main 
ring as shown in this figure. The size of the figure is 
$1.8 \times 1.8$~mm.}
\label{ringtrap}
\end{center}
\end{figure}

Examples of the "magnetic trap" which is formed near the $B=0$ point are 
shown in Fig.s~\ref{feeding} and \ref{ringtrap}. When sufficient number of 
atoms is fed into the MOT, and the MOT is approximately spherically symmetric,
the MOT has a bright spot in the center, and 
is accompanyed by a wing whose intensity decreases roughly
inverse to the distance. The size of the MOT generally increases as the 
detuning of the trapping laser is increased, and its pattern tends to form
a ring shape. Even in that situation the "magnetic trap" is frequently observed
near the "B=0" point which does not necessarily overlap with the main
ring as seen Fig.~\ref{ringtrap}.

In the next section we describe briefly general theoretical
considerations on the beam source.
In the third section we describe the configuration of our atomic beam 
apparatus in detail. Then, the experimental results of the atomic trap 
and beam are given in the following section.
A few demonstrative interferometric experiments
are given in the last section.


\section {Theoretical consideration}
\label{theory}


Unlike an optical beam, an atomic beam 
suffers loss of coherence caused by collisions 
between atoms in the beam. If the diameter
of the source is $d$, and average atomic velocity is $v$, the atom stays in the
source at least $t \geq d/v$.
If $v=0$, it is the time that the gravity pull out the atom from the 
source, $t \geq \sqrt{2d/g}$, where $g$ is gravitational
acceleration, $g=9.8$~m/s. The atom has to escape from the source before 
the coherence is lost by collisions between trapped atoms. Therefore,
\begin{equation}
t \leq \frac{1}{\alpha n},
\end{equation}
where $\alpha$ is the two-body collision rate, and $n$ is the density of the
atom. Then, the maximum density of the atom is
\begin{equation}
n \leq \frac{v}{\alpha d} \mbox{     or     } \sqrt{\frac{g}{2d\alpha^2}},
\end{equation}
and the maximum flow of the atomic beam $N$ is,
\begin{equation}
 N=\frac{nd^3}{t}=\frac{v^2d}{\alpha} \mbox{     or     } \frac{gd^2}{2\alpha}.
\end{equation}
The flux of the atomic beam decreases with $d$.
However, the brightness increases, because it is propotional to $d^2$. 
The flux decreases also with the kinetic energy of the atom. The brightness
, however, is constant.

For atoms with a large collision cross sections such as metastable rare 
gases, the maximum flux $N$ is not large. For 1s$_5$ 
metastable neon the two-body loss rate is
$\alpha=2.5 \times 10^{-10}$~cm$^3$s$^{-1}$\cite{neondecay}.
The typical values of our source in the best experimental conditions are 
$v=20$~cm~s$^{-1}$ and
$d=60$~$\mu$m. Then, the ultimate maximum density is between 
$10^{13}$~cm$^{-3}$ and $10^{12}$~cm$^{-3}$,
 and the atomic flow $10^{10}$~s$^{-1}$ to $10^8$~s$^{-1}$.

In a commonly used MOT, $d$ is much larger than $d=60$~$\mu$m.
Simple MOT theory tells that the trap size is determined 
by the surface where
atoms in the high-field-seeking magnetic state is resonant to the red-detuned
trapping laser. Atoms are assumed to stay dominantly in the high-field-seeking 
state. This is correct as an approximate physics of MOT, 
but in actual experiment the situation is much more complex. 
To keep atoms always in the high-field seeking states, the magnetic field and 
the k-vector of the optical field must be parallel. This cannot be done 
in three-dimensional quadrupole field. Therefore, Atom has probability to be
pumped to a low-field seeking state, and this is spatially dependent.
The intensity of the laser also infuences the MOT. If the laser intensity 
is weak, the atom interacts with the laser only at a large distance from the
center of the magnetic quadrupole field. When the intensity is strong, 
the atom interacts even near the center where there is no preferentially 
populated magnetic sublevels. 
The MOT is generally very stable, and can
resist against intensity and spatial imbalance of the trapping laser beams.
This produces various structure of atomic cloud which is sometimes 
difficult to control.
Those factors, we believe, creates the bright spot around $B=0$ point.

Experimentally, when
the laser intensity and magnetic field gradient are large, and the laser 
detuning is large, a small bright spot of very cold atoms near the $B=0$
point is commonly observed. This is the atomic source we use to generate
cold atomic beam, and we call his source as "magnetic trap"
in the following sections, though the bright spot may not be 
the standard magnetic quadrupole trap which works without optical field.

The path of atoms that feed the trap also influences the maximum beam
intensity. If the atom moves along a line keeping its diameter,
collision rate during the approach is equal to the rate in the trap.
Therefore, the maximum beam 
density is reduced by a ratio between the path length and $d$.
If atoms approach from various direction on a one-dimensional line,
the atomic density decreases as $1/r$, where $r$ is the distance from
the center. The total loss by binary collision is only logarithmically 
divergent. If atoms approach two dimensionally, the loss increases
only a factor two. The dynamics can be estimated from the fluorescence
pattern of the trapped atoms. If atoms approach two dimensionally,
the fluorescence intensity is inversely proportional to $r$ provided
that their velocity is constant.

\section {Neon beam chamber}


\begin{figure}[htbp]
\begin{center}
\includegraphics[width=12cm]{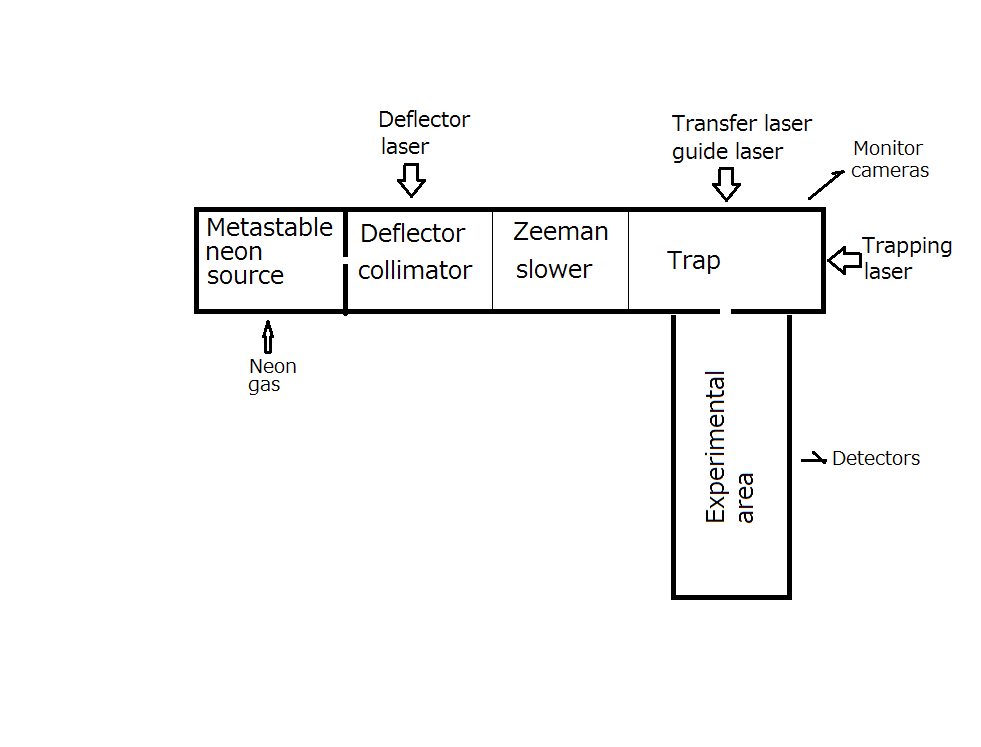}
\caption{Schematics of the experimental vacuum system. Metastable neon atoms
are generated in the source section, travel right, and are
trapped in the trap section. Trapped atoms 
are illuminated by the transfer laser,
released from the trap, and fall into the experimental area which is
located below the trap section.
}
\label{schematics}
\end{center}
\end{figure}

\begin{figure}[htbp]
\begin{center}
\includegraphics[width=8cm]{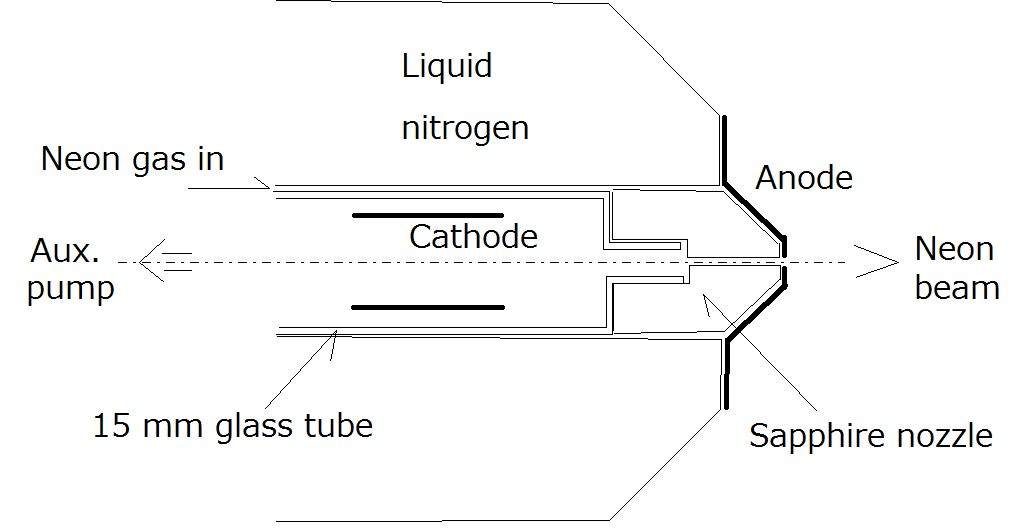}
\caption{Metastable neon atoms are generated by weak discharge
in the sapphire capillary which is cooled by liquid nitrogen.
The diameter of the capillary is 0.5~mm, and its length 
approximately 10~mm. Discharge current is typically 10~mA.}
\label{nozzle}
\end{center}
\end{figure}

\begin{figure}[htbp]
\begin{center}
\includegraphics[width=8cm]{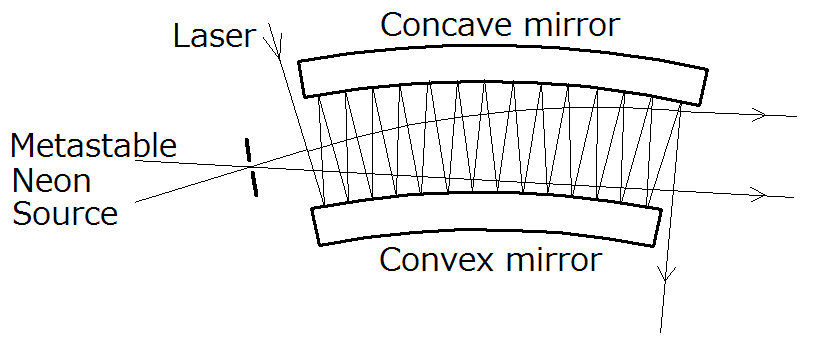}
\caption{The deflector-collimator. The figure shows one of two 
mirror pairs. 
The concave and convex mirrors
are placed concentrically. The radii of the 
curvature are 1.38 and 1.30 meters, respectively. The mirrors are rectangular 
with the size of $40 \times 150$~mm. Another pair of mirrors are placed 
perpendicular to this pair.
The laser beam reflects between two mirrors typically 30 times.}
\label{deflector}
\end{center}
\end{figure}

\begin{figure}[htbp]
\begin{center}
\includegraphics[width=8cm]{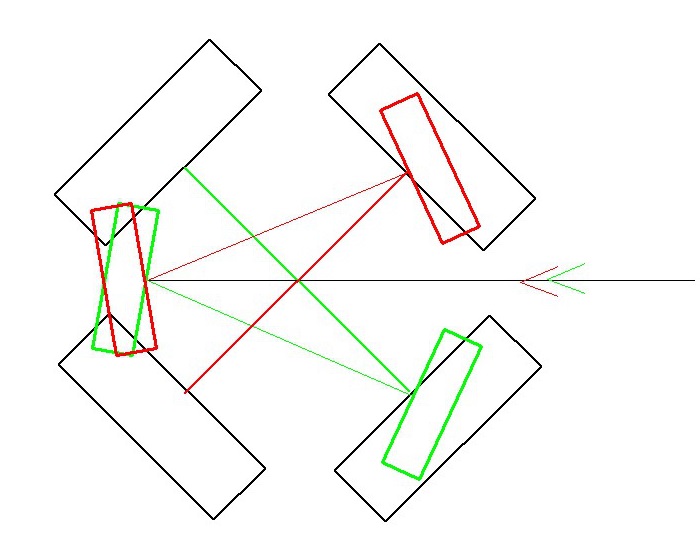}
\caption{The input mirror configuration of the deflector. 
The input laser beam is focused
at approximately 1.3~m upstream of the inlet. The beam diameter 
expands to
approximately 20~mm at the inlet. The beam is divided into two
by two small rectangular flat mirrors, and distributed to each 
deflector mirror pair (bigger rectangles in the figure). 
The configuration of the input mirrors 
equalizes the path length
from the pre-focus point to the deflector mirror pairs.
The size and pattern of the laser are kept approximately same
when the laser travels zigzag between two mirrors.}
\label{input}
\end{center}
\end{figure}

\begin{figure}[htbp]
\begin{center}
\includegraphics[width=12cm]{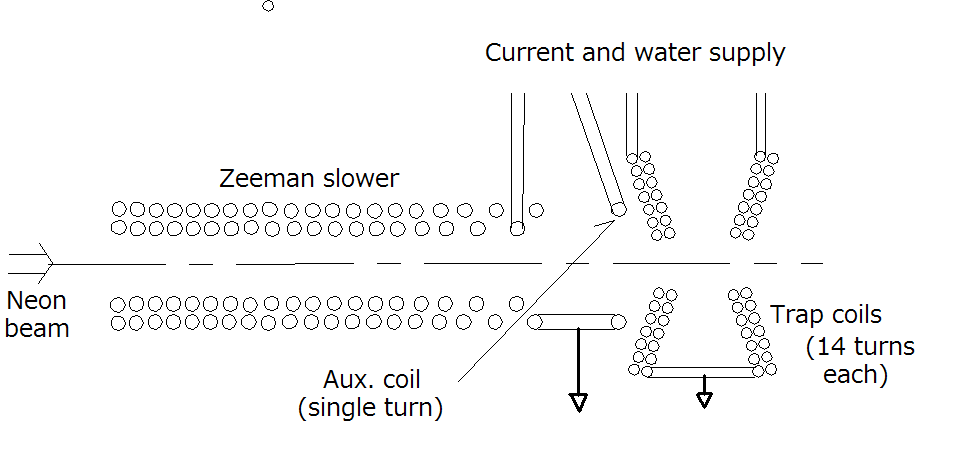}
\caption{The metastable neon beam enters in the Zeeman coil made of two 
layers of
3/8 inch copper tube, then, moves to the trap coils which produces a
quadrupole magnetic field. A single-turn coil is placed between 
the Zeeman and trap coils to optimize the beam transfer into the trap. 
The trap coils 
are made of 1/8 inch copper tube, The shape of the trap coil is a truncated 
cone with the apex angle of 142 degrees to allow the trapping laser beams
into the center of the quadrupole magnetic field.
The current of the Zeeman coil is typically 190~A, The trap coils produce
the magnetic field gradient ot 1=T/m along the beam axis at 100~A.
All coils are cooled by flowing pressurized water typically at 1.7~MPa.}
\label{zeemancoil}
\end{center}
\end{figure}

\begin{figure}[htbp]
\begin{center}
\includegraphics[width=8cm]{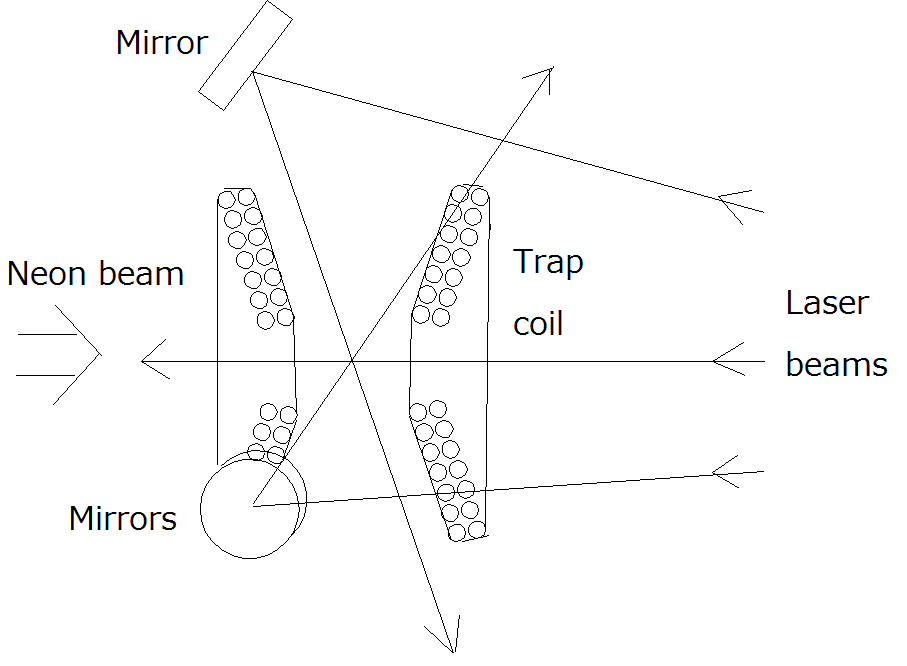}
\caption{Trap coils. 
Trap coils are made of 1/8 inch copper tube, and are cooled by pressurized
water flow at 1.7~MPa. The figure show top view. The gap between
two coils is approximately 4~cm. They produce the magnetic field
grating of 1~T/m (along the symmetry axis) with the current of 100~A.}
\label{trap}
\end{center}
\end{figure}

\begin{figure}[htbp]
\begin{center}
\includegraphics[width=12cm]{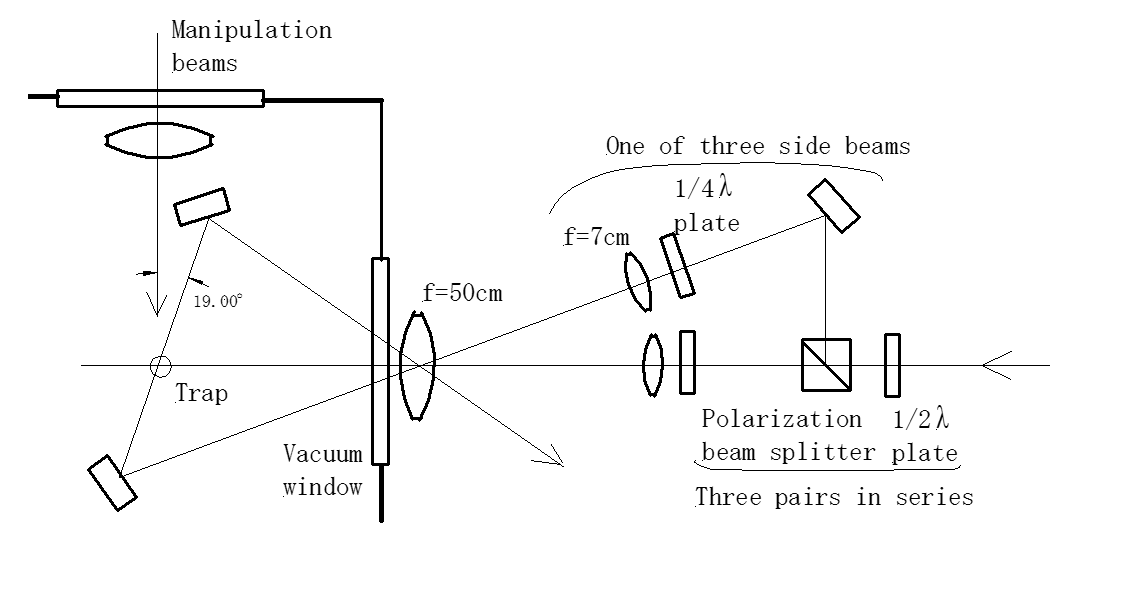}
\caption{Optical arrangement of the trapping laser beams. The trap laser beam
is divided into four beams by three pairs of a half-wave plate and a
polarization beam splitter. The three pairs are 
placed in series along the symmetry axis. Then, four laser
beams are circularly polarized of the same polarity, 
are expanded their rasdius 
with two lenses, and led into the trap vacuum chamber, three side beams are
directed towards the trap from the backside, 
and after passing through the trap the second mirror
directs the beams outside the vacuum chamber to avoid the laser beams 
hitting the vacuum chamber wall.}
\label{optics}
\end{center}
\end{figure}

\begin{figure}[htbp]
\begin{center}
\includegraphics[width=8cm]{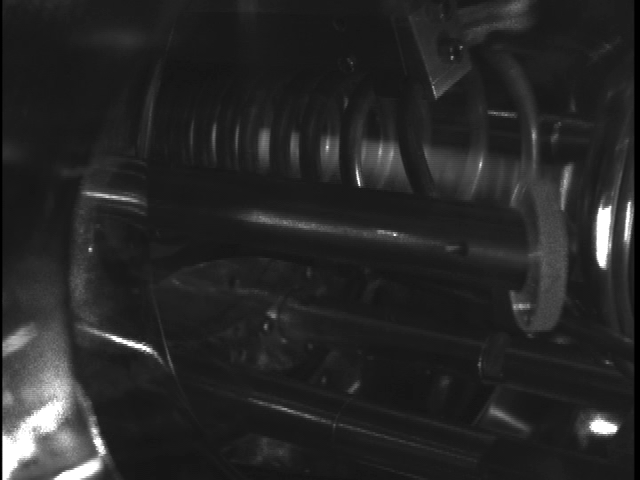}
\caption{A part of the Zeeman slower. 
When the deflector-collimator is working, the fluorescence from 
the metastable neon atoms should be clearly visible.}
\label{slowerfigure}
\end{center}
\end{figure}

\begin{figure}[htbp]
\begin{center}
\includegraphics[width=8cm]{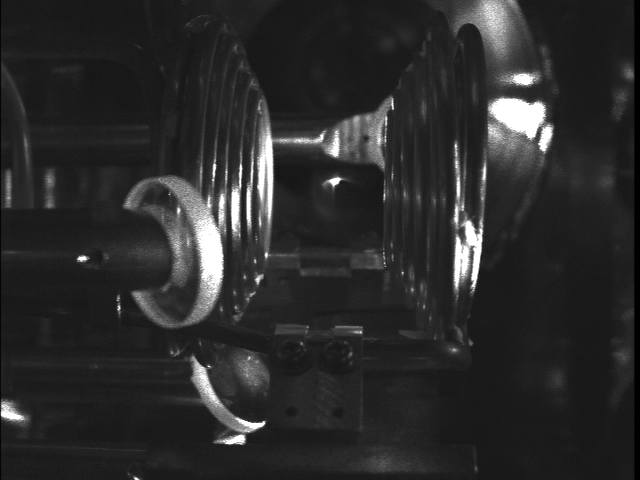}
\caption{Photo of the trap coils, trapped atoms and a part of 
the manipulating mirrors for the trapping laser. Atomc come from left.}
\label{trapfigure}
\end{center}
\end{figure}

\begin{figure}[htbp]
\begin{center}
\includegraphics[width=8cm]{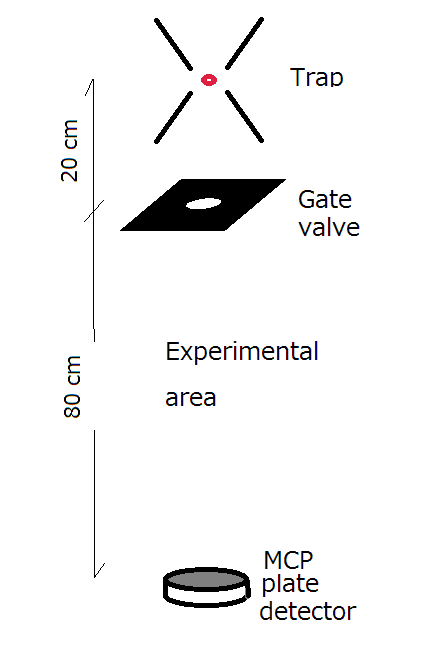}
\caption{The experimental area.
Slow 1s$_3$ metastable neon atoms fall vertically. They pass
through a racetrack shape gate valve of $ 5 \times 8$~mm which separates
the experimental area at
20~cm below the trap.
Atoms are detected by a micro-channel detector (MCP)
equipped with fluorescent plate, which is placed typically
80~cm below the gate valve. The pressure in the experimental
chamber is typically $5 \times 10^{-8}$~Pa. }
\label{area}
\end{center}
\end{figure}


The vacuum chamber consists of a metastable beam source, a 
deflector-collimator, 
a graded Zeeman magnetic 
field coil to slow metastable atoms, and the MOT
region.  They are placed horizontally. The experimental chamber is attached 
vertically below the MOT section. (See Fig.~\ref{schematics}.)

The metastable neon atoms are generated by a weak discharge through
the sapphire capillary. Its diameter was 0.5~mm, and the length is
5~mm. The sapphire is cooled by liquid nitrogen on contact.
(See Fig~\ref{nozzle}.) This section is pumped by a 800 litter turbo pump
and is connected to the downstream through a 3~mm$\phi$ hole.
The neon atoms are led into the deflector-collimator section through
this hole. (See Fig~\ref{deflector}.) 

The collimator consists of two pairs of concave-convex
mirrors. Two pairs are placed diagonally so that each pair collimates
the two orthogonal radial directions of the beam. 
The curvature of the mirror
are 1.30~m for the convex mirror and 1.38~m for the concave mirror.
Two mirrors are placed concentrically. 
The deflector laser at 640.4~nm is focused at 1.3~m upstream of the 
inlet of the collimator sets, and is expanded to approximately 20~mm
in diameter at the inlet. It is divided into two, and each half is
led into two mirror pairs, respectively. Four rectangular mirrors are
used to feed the laser into two mirror pairs 
keeping the same optical path length. 
The laser travels zigzag between two mirrors.
The number of reflection is typically 30, which can be changed by 
adjusting the input angle of the laser. 
The pattern of the laser beam on the mirror
is kept approximately same because of the concentric arrangement of 
convex-concave mirrors and pre-focusing of the laser beam.
The system collimates approximately 0.1 radian of the metastable neon beam.
This results enhancement of the trapped neon atoms typically
 by a factor of 50. The frequency stability of the laser required for
the efficient collimation was approximately 10~MHz.

Neon atoms are decelerated inside the Zeeman coil which is made of two
layers of 3/16 inch copper tube. The length of the coil is approximately 
45~cm, and the coil produces axial magnetic field which is maximum at the 
inlet and decreases to zero
at the exit of the coil. (See Fig.~\ref{zeemancoil}.) Atoms are then led 
into the trap coils which produces quadrupole magnetic field with the
symmetry axis parallel to the atomic beam direction. Since we use relatively 
high field gradient for the MOT, 
a single turn coil is inserted between the Zeeman coil 
and the trap coil to improve the coupling efficiently of atoms into the 
trap region. The magnetic field gradient was between 1.5~T/m and 0.4~T/m. 
All coils are cooled by pressurized water through the copper tube at the 
pressure of 1.7~ MPa. (See Fig.~\ref{trap}).

Figure~\ref{slowerfigure} shows a part of the Zeeman coil. Fluorescence from
the trapping laser should be clearly visible when the 
deflector-collimator is operating properly.
Figure~\ref{trapfigure} is the trap part. The diameter of the
mirrors to guide the side laser beams is 19~mm. Typical pressure 
in the trap region is $3 \times 10^{-7}$~Pa, when the atomic beam is on.

MOT was achieved by using four laser beams\cite{tetra}. 
One laser beam heads towards the beam source, 
which is parallel to the atomic beam. The remaining
three laser beams are shoot at approximately 71 degrees to the atomic 
beam axis.
The optics to distribute trapping beams are shown in Fig.~\ref{optics}. 
The laser beam at 640~nm from the Coherent 699 dye laser is split 
into four by using four pairs of
half-wave plate and  polarization beam splitter. Then, the polarization of
each beam is converted to
circular of the same polarity 
with a 1/4 wave plate. Then, they are expanded with two lenses 
before they are led into the vacuum chamber. 
The maximum power density of a single beam was approximately 90~mW/cm$^2$.
The transfer laser beam at 598~nm is focused in the trap from the top.
This laser transfers trapped neon atoms 
in the 1s$_5$ state to the 1s$_3$ state by optical pumping and releases atoms
from the trap. The released atoms fall vertically into the experimental area.

The experimental area is separated from the rest of the vacuum chamber through
a race-track shaped gate valve. This part is a vertically placed 76~cm long
6 inch tube with several ports to insert parts necessary for experiment. 
Metastable atoms are detected by a micro-channel plate
detectors (MCPs) equipped either fluorescence plate or an anode. 
Typical pressure in
this region was $3 \times 10^{-7}$~Pa. (See Fig.~\ref{area}.)

\section{Experimental results of the trap}


\begin{figure}[htbp]
\begin{center}
\includegraphics[width=8cm]{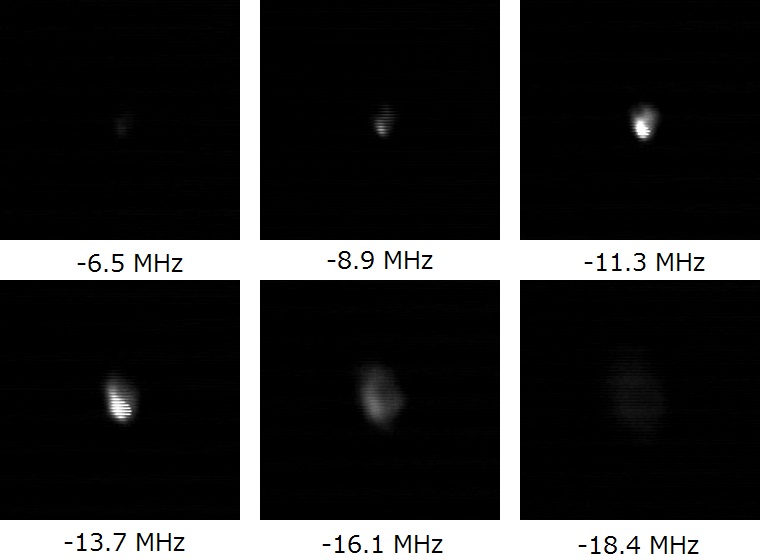}
\caption{The trap pattern as a function of the detuning of 
the trapping laser, when operating parameters are in the commonly used
range. Laser intensity of each arm is 4~mW/cm$^2$. Numbers below each
figure is the detuning of the trapping laser.
Natural line width of
the trap transition, 1s$_5$ - 2p$_9$, is 4~MHz.
The size of each picture is $4.6 \times 4.6$~mm.
}
\label{fig17_4}
\end{center}
\end{figure}

\begin{figure}[htbp]
\begin{center}
\includegraphics[width=8cm]{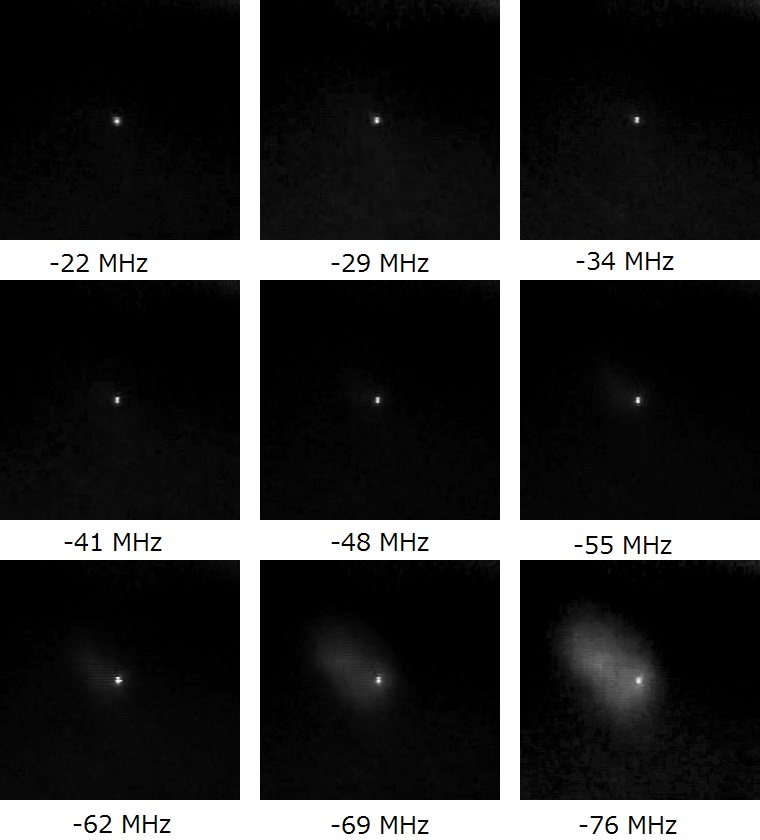}
\caption{The trap pattern as a function of the detuning of the trapping laser.
The number below the figure is the detuning of the trapping laser.
For those figures, the deflector-collimator is not in use. 
Therefore, the source
intensity is relatively weak.
Bright spot near the zero magnetic field persists over 
a wide range of detuning.
The brightness of the picture is normalized independently in each picture.
The size of each picture is $4.6 \times 4.6$~mm.}
\label{trapweak}
\end{center}
\end{figure}

\begin{figure}[htbp]
\begin{center}
\includegraphics[width=8cm]{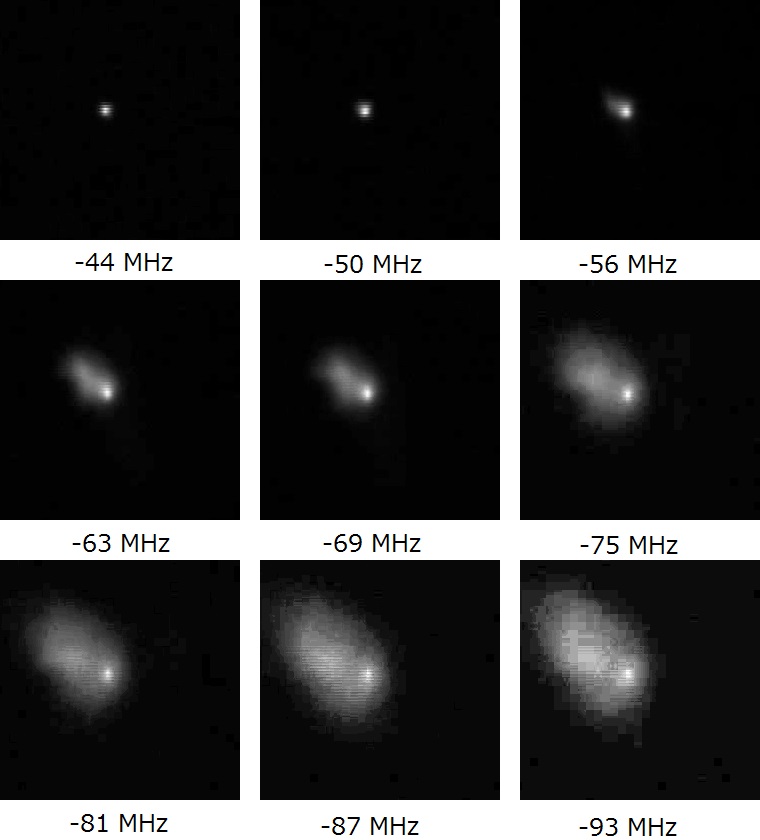}
\caption{The trap pattern as a function of the detuning of the trapping laser.
The number below the figure is the detuning of the trapping laser.
Neon beam is collimated after the nozzle. The fluorescence intensity is nearly
100 times stronger than in Fig.~\ref{trapweak}. The bright spot persists, and
it is accompanied by big cloud around the spot. This is due to a shorter 
lifetime of the bright spot due to high atomic density.
The brightness of the picture is normalized independently in each picture.
The size of each picture is $4.6 \times 4.6$~mm.}
\label{trapstrong}
\end{center}
\end{figure}

\begin{figure}[htbp]
\begin{center}
\includegraphics[width=12cm]{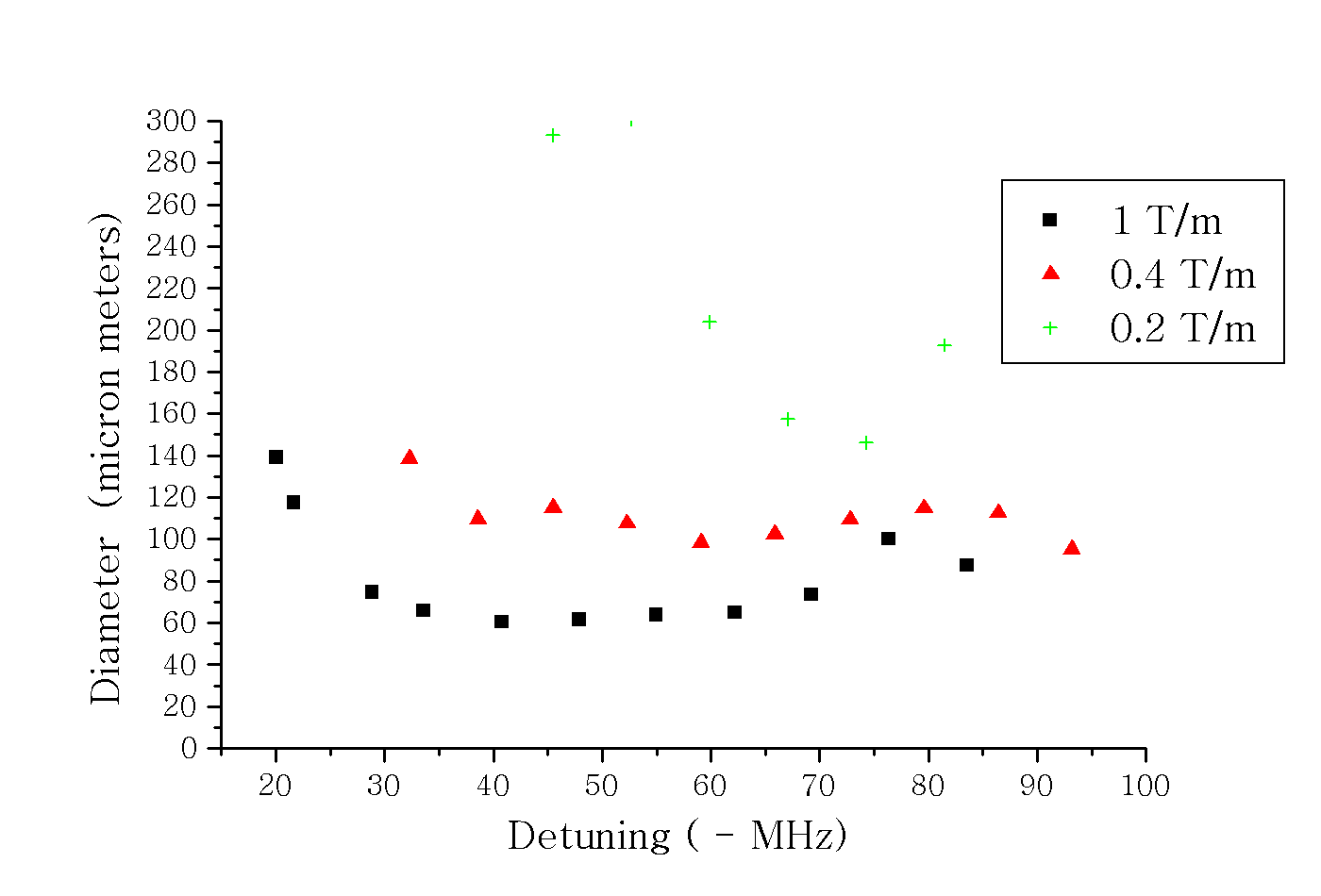}
\caption{The trap diameter as a function of the detuning of 
the trapping laser for three quadrupole field intensities.
The source intensity is weak. Deflector-collimator is not in operation.}
\label{widthweak}
\end{center}
\end{figure}

\begin{figure}[htbp]
\begin{center}
\includegraphics[width=12cm]{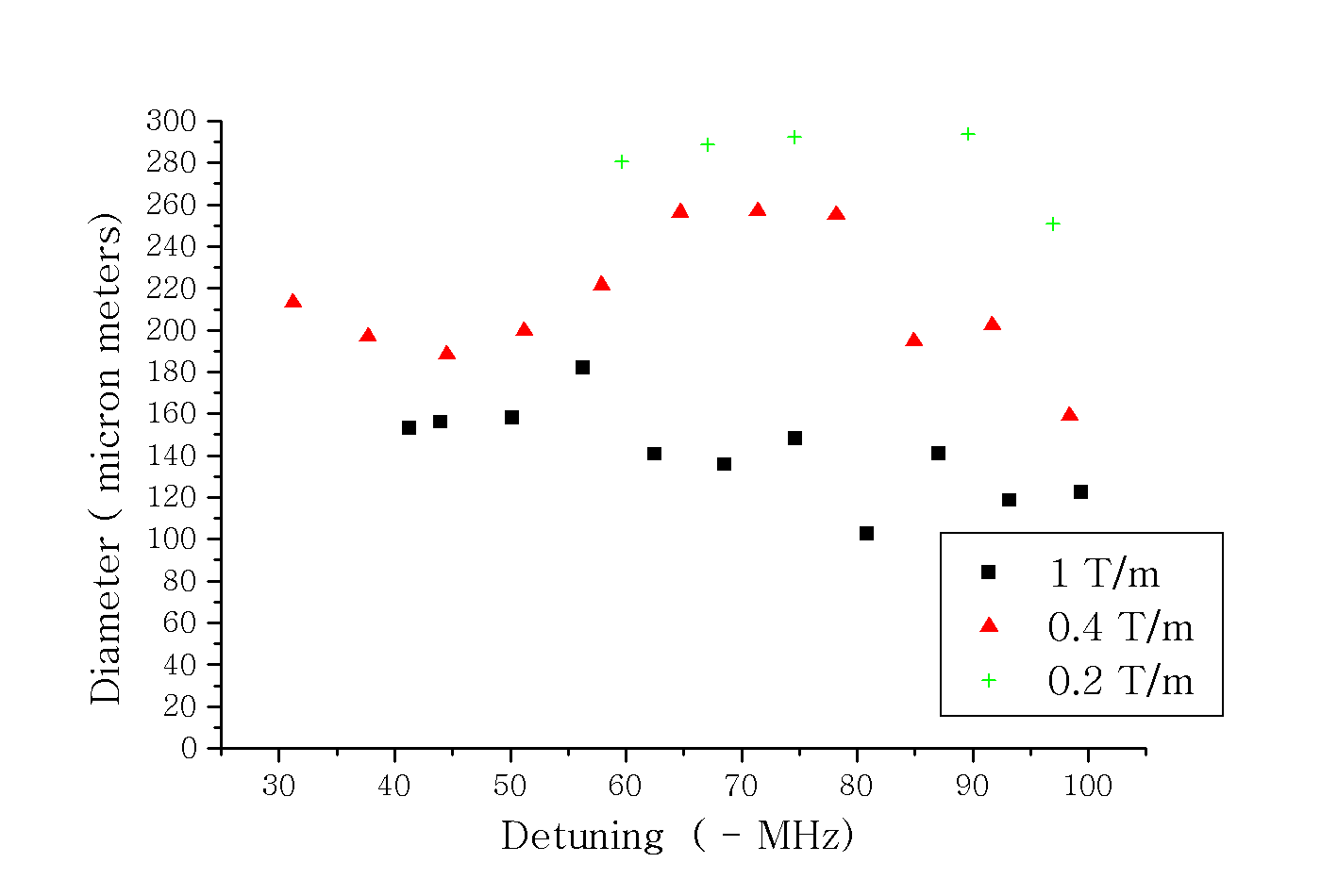}
\caption{The trap diameter as a function of the detuning of the trapping laser
for three quadrupole field intensities.
The source intensity is strong. Deflector-collimator is in operation.}
\label{widthstrong}
\end{center}
\end{figure}

\begin{figure}[htbp]
\begin{center}
\includegraphics[width=8cm]{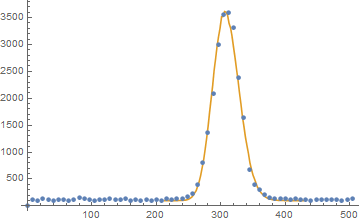}
\caption{Typical time-of-flight pattern measured with a MCP
placed 46~cm below the trap, Abscissa is the time-of-flight from the
source to the detector in ms. Ordinate is 100 times of the
 atom counts in each bin (8~ms wide). The fitting curve is for the 20~cm/s
gaussian velocity distribution at the source.}
\label{feb28tof}
\end{center}
\end{figure}

\begin{figure}[htbp]
\begin{center}
\includegraphics[width=8cm]{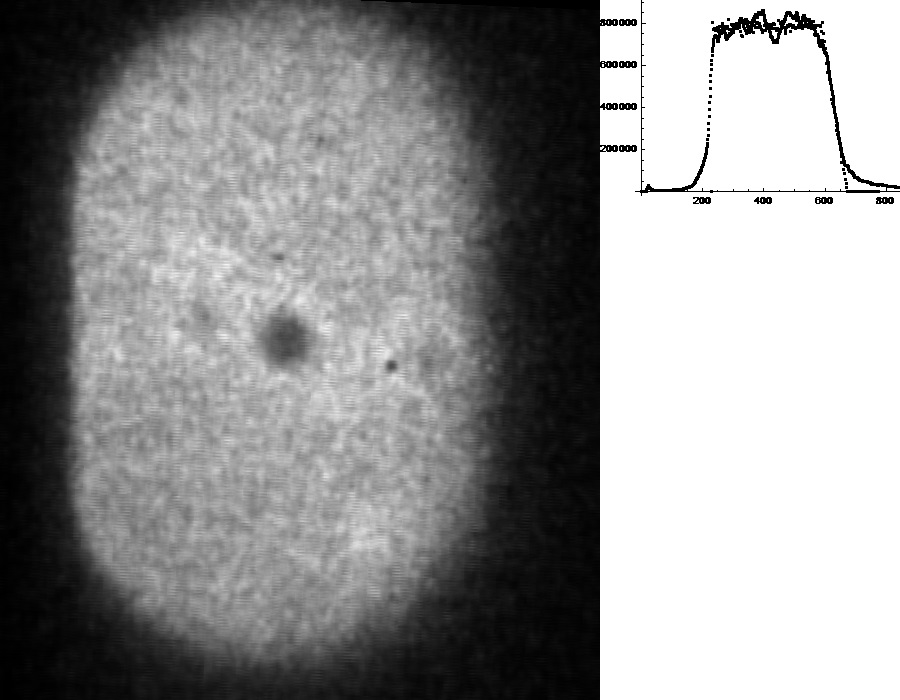}
\caption{The image of the gate on the micro-channel plate detector(MCP)
which is placed 94~cm below the trap. 
The curve on the right shows horizontal intensity variation.
The trap is located near the left edge of the race-track 
pattern. Therefore, the velocity distribution of the trapped
atoms can be estimated from the slope of the right edge.
Gray dots shows the curve which is obtained  by assuming
gaussian distribution of $v=\sqrt{3} \times 30$~cm/s.
The exposure time of the figure is 3.3~s.
Dark spot near the center is the defect of the MCP.
The horizontal size of the picture is 13.5~mm.}
\label{gateimage}
\end{center}
\end{figure}

\begin{figure}[htbp]
\begin{center}
\includegraphics[width=8cm]{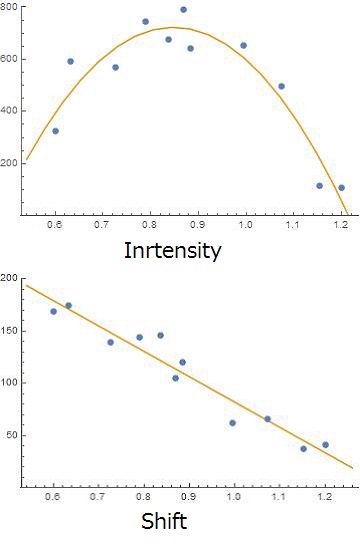}
\caption{Spatial stability of the trapped atoms.
The influence on the trap pattern 
by the imbalance of trapping laser beams.
In this graph, the intensity of three side beams are kept 
approximately equal, and
the intensity of the on axis beam is changed. Abscissa is the 
intensity ratio between the on-axis and side beams. 
The ordinate of the lower figure is the relative shift of the
trap position in $\mu$m.
The trap position is seen to
move roughly the bright spot diameter. }
\label{feb19shift}
\end{center}
\end{figure}


We operated the MOT at large magnetic field gradient, high laser intensity, 
and high laser detuning. The magnetic field gradient was typically 1T/m, and 
was varied from 
0.2~T/m to 1.5~T/m. The laser power density was from 90~mW/cm$^2$ to
20~mW/cm$^2$, and the detuning up to -120~MHz. Under this operating condition
the MOT produced an additional stable trapping region near zero magnetic
field point. The temperature of this $B=0$ trap was usually lower 
than the Doppler limit temperature\cite{chu}, and had a small size.

For the standard operating condition of MOT the trap size is expected
to increase linearly with the detuning of the trapping laser. Atoms in the
high magnetic field seeking state
are reflected back at the ellipsoid surface resonant to
the cyclic transition 1s$_5$ to 2p$9$. The ellipsoid expands linearly with
the detuning.
Figure~\ref{fig17_4} shows typical MOT pattern of the trapped atoms with
the laser power density of a few times of the saturation intensity and 
the detuning of a few times of the natural line width. Trap size is seen to
increase  with the detuning of the laser as expected from this
simple theory.

The situation is different when both the laser power and detuning
are large, and the magnetic field gradient is large. 
(See Fig.s~\ref{trapweak} and 
\ref{trapstrong}.) We were always able to generate a bright small spot 
near the $B=0$ point by adjusting the trap laser direction and intensity.
Figures~\ref{trapweak} and \ref{trapstrong} show trap patterns as a function
of laser detuning for weak and strong metastable source intensity, 
respectively. In
Fig.~\ref{trapstrong} the source intensity is approximately 50 times stronger
than in Fig.~\ref{trapweak}. In each picture the peak intensity rightness is
normalized for better visibility of the atomic density profiles.
A bright spot around the $B=0$ point persists up to
the detuning of -100~MHz.  The field gradient along the symmetry axis is 
1~T/m, and the laser intensity of 80~mW/cm$^2$ for each beam.
The minimum diameter is approximately 60~$\mu$m for the weak source 
(Fig.~\ref{trapweak}), and 100~$\mu$m for the strong source 
(Fig.~\ref{trapstrong}). As shown in 
Fig.~\ref{widthweak} and \ref{widthstrong} the diameter
of the trapped atoms do not change much over a wide range of laser detuning. 
It 
increases with decreasing field gradient. However, the variation is slower 
than the inverse law. The diameter is larger when the source intensity is
strong. This indicates that the temperature of the trap decreases slightly
when the field gradient is lower. It is probably hotter when the source
intensity is stronger, because atoms stay shorter time in or near 
the trap  before removed by inelastic collisions.

The trapped atoms are released by the transfer laser 
at 598~nm focused on the bright
$B=0$ spot. The diameter of the laser at the focal point was approximately
30~$\mu$m, and the divergence angle between 1/5 and 1/10. The optimal intensity
of the transfer laser varies with the atomic density of the bright spot  
reflecting the collisional lifetime. When the feeding of the atom 
into the trap was large as shown in Fig.~\ref{trapstrong}, 
the generated $1s_3$ atomic beam was
strongest when the transfer laser was several$\mu$W. When the feeding was weak 
as in Fig.~\ref{trapweak}, the optimal intensity was in the order of 100~nW.
If the intensity of the transfer laser was not excessive, the $1s_3$ atomic 
beam drew a clear shadow image of the gate vlve on the MCP placed at the
bottom of the eperimental chamber as shown in Fig.~\ref{gateimage}.
We can estimate the atomic velocity distribution at the source.
In this figure the horizontal position of
the trap was placed close to 
the left edge of the gate image. Therefore, the slope of the right edge
reflects the velocity distribution. The typical velocity obtained by
this method is 50~cm/s resulting the temperature of 260~$\mu$K. This method 
gives higher temperature than the direct measurement described below, 
probably because it
includes the broadening of the edge by the limited resolution of an atom image, the trap size, and other disturbance during the flight. 

More accurate measurement was obntained by the 
time-of-flight measurement (Fig.~\ref{feb28tof}). 
A short transfer laser (2~ms) 
illuminated the trap, and the time-of-flight of the released 1s$_3$ atoms was
detected by a MCP placed at 46~cm below the trap. Typical one-dimensional
velocity was 20~cm/s to 30~cm/s, 
corresponding to the temperature of 45 to 100~$\mu$K.
Since the transfer laser was illuminated for a short time 
and with low-repetition rate, this value shows 
the temperature of the steady state, where trapped atoms are lost only 
by inter-atomic collisions.

Another method is to estimate from the diameter of the bright spot. 
Assuming that this bright spot is practically the magnetic 
quadrupole trap, its radius can be calculated from the
Zeeman shift. The typical radius is 60~$\mu$m at 1~T/m.
The temperature is then 30~$\mu$K, if the atom is in the $M=1$ level, and
60~$\mu$K, if in the $M=2$ level of the 1s$_5$ state. This result, 
suggests that the attractive Zeeman force of the quadrutic magnetic field
plays an important role for the forming of the bright spot neaar $B=0$ point.

To test the spatial stability of the bright spot 
we measured the intensity and the
position of the trap when the relative intensity between the on-axis laser beam
and the side beams were varied. Figure~\ref{feb19shift} shows the result when 
the intensity of the three side beams were kept approximately equal, and the
intensity of the on-axis beam was varied. The trap persisted 
even when the on-axis
beam intensity was varied more than a factor two. In addition the shift of the 
trap position was in the same order of 
the trap diameter over the entire intensity range.
Furthermore, the trap intensity was maximum when the on-axis laser intensity 
was slightly less than that of the side beams. At exactly $B=0$ the force 
excerted on the atom is that of the molasses\cite{chu}. Atoms are not trapped.
However, if the favorable  condition to the low-field-seeking levels
is formed as a result of intensity imbalance, atoms will be trapped as in
the case of the magnetic quadrupole trap. 

A relable way to estimate the atomic density in the "magnetic trap" is to
analyze the recovery curve of the density of the trapped atoms 
when the transfer laser is
turned off. The recovery curve is given by
\begin{equation}
n(t)=n_f\frac{A\exp{t/\tau}-1}{A\exp{t/\tau}+1}
\end{equation}
with $A=(n_f-n_i)/(n_f+n_i)$ and $\tau=1/(\alpha n_f)$, 
where $n_f$ is the steady state intensity,
$n_i$ is the density when the transfer laser is on, and $\alpha$ is
the binary collision rate. For our case 
$\alpha =2.5 \times 10^{-10}$cm$^3$s$^{-1}$\cite{neondecay}.
We obtained the shortest recovery time of $\tau = 10^{-3}$~s.
This gives $n=4 \times 10^{12}$cm$^{-3}$. This value is not far from
the maximum density we can expect, whch is described in Sec.~\ref{theory}.

\section{Numerical simulation}


\begin{figure}[htbp]
\begin{center}
\includegraphics[width=8cm]{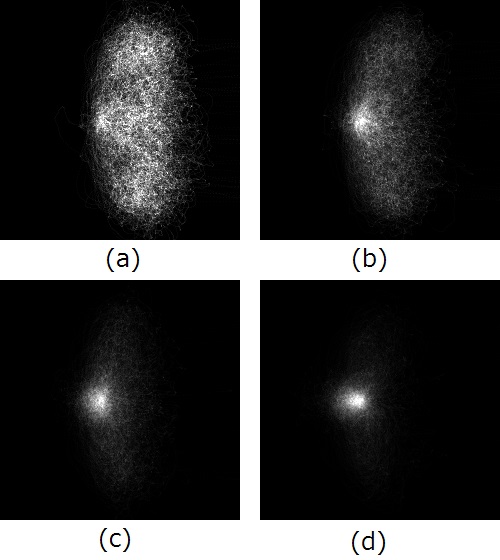}
\caption{Trap pattern of four-beam MOT. Four figures show the change of the
trapped atom pattern when one of the trapping laser intensity is slightly
changed.
The picture show the pattern accumulated from the trajectory of 
50 atoms.
(a) The intensity of the four laser 
beams are equal and 60 times of the saturation intensity. (b) The axial beam 
is 0.95 times of the other beams. (c) 0.9 times. (d) 0.85 times.
The detuning of the laser is 10 times of the natural width.
The magnetic field gradient is 10~mT/cm along the atomic beam direction.
Atoms fly from right to left. The piture size is $2 \times 2$~mm.}
\label{simulation}
\end{center}
\end{figure}

\begin{figure}[htbp]
\begin{center}
\includegraphics[width=10cm]{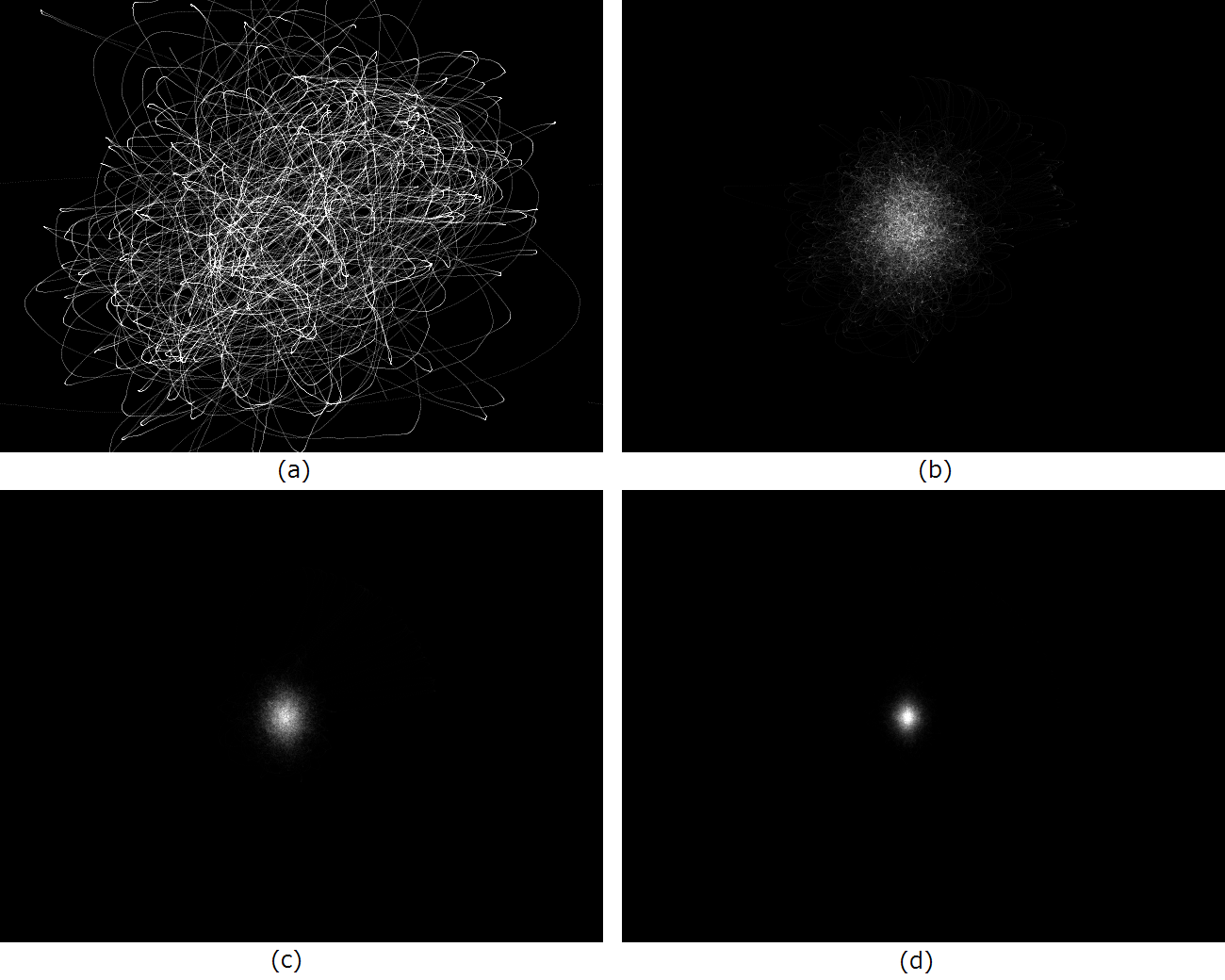}
\caption{Trap pattern of the standard six-beam MOT. 
Four figures show the change of the
trpped atom pattern with the laser intensity.
The picture show the pattern accumulated from the trajectory of 
50 atoms.
The intensity of the  laser 
beams is (a) three times of the saturation intensity
 (b) 10 times (c) 30 times (d) 100 times. The lowest
power to form trap is a little less than three times of the saturation
intensity..
The detuning of the laser is 10 times of the natural width.
The magnetic field gradient is 10~mT/cm along the atomic beam direction.
Atoms fly from right to left. The piture size is $3 \times 4$~mm.}
\label{simulation6}
\end{center}
\end{figure}

\begin{figure}[htbp]
\begin{center}
\includegraphics[width=8cm]{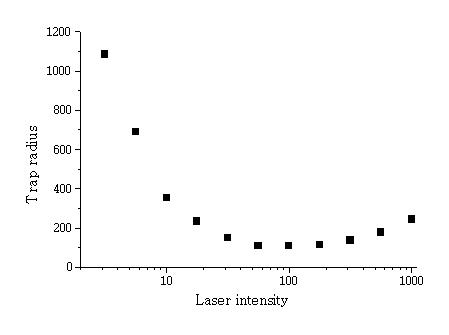}
\caption{The trap radius vs trap laser intensity. The abscissa is in unit of
the saturation power. The ordinate is in unit of $\mu$m. The magnetic field
gradient is 1~T/m. The laser detuning is 10 times of the saturation intensity.}
\label{radius} 
\end{center}
\end{figure}

\begin{figure}[htbp]
\begin{center}
\includegraphics[width=8cm]{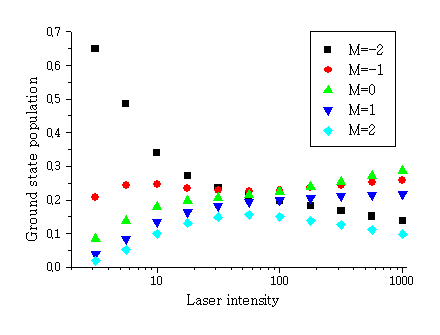}
\caption{Distribution among magnetic sublevels in the lower state 1s$_5$.
The abscissa is in unit of the saturation power.The magnetic field
gradient is 1~T/m. The laser detuning is 10 times of the saturation intensity.}
\label{population}
\end{center}
\end{figure}


We tried to simulate the experimental results in the
previous section by semi-classical numerical
simulation. The result justities only qualitatively our 
experimental observations
that atoms tend to accumulate near $B=0$ point
under some conditions.
The calculation uses a simplified semiclassical method in which we assume (1)
the atom has definite location and velocity, (2) it moves on the specific
quantum state potential, and (3) the atom repeats the
following cycle. The atom is initially in the ground state. Then, its
upper-state probability increases by the interaction with 
trapping laser field. When the elapsed
time
becomes equal to the natural lifetime of the excited state, the atom decays
to the ground state by emitting a spontaneous photon, The time evolution
of the excited state poplulation is assumed to be
\begin{equation}
|\psi_i(t)|^2=a\{1-\exp(-bt)\}^2,
\end{equation}
where $a$ and $b$ are constants determined by the conditions 
in which $|\psi_i|^2$ is accurate at 
small $t$ and approaches to the stationary value at large $t$.

Figure~\ref{simulation} shows an example of the trapped atomic cloud with
a four-beam MOT. Four pictures in the figure show the change of the
trap pattern as the on-axis laser intensity is changed.
When the intensity of all lasers are equal, the trap is large and has 
elliptical shape. When the intensity of the on-axis laser is slightly
reduced relative to the other lasers, atoms start to be accumulated
around the $B=0$ point. As the on-axis intensity is further reduced, 
the bright spot persists and moves slightly towards downstream. 
This behavior is qualitatively
the same as the experimental observation. However, the stable range of
the bright spot is less than that observed in the experiment. 

Though we do not have experimental verification, the four laser beam 
configuration is probably not the necessary condition to produce the
$B=0$ spot. Figure~\ref{simulation6} shows the trap pattern when the 
laser intensity is changed from near threshold to 100 times of 
the saturation intensity for the standard six-beam MOT.
Together with Fig.~\ref{radius} the influence of the laser intensity
is dramatic. The diameter shrinks as the laser intensity increases
towards the saturation intensity at the $B=0$ point. 
It is clear from Fig.~\ref{population} that the population is
more opr less equally populated among all magneetic sublevels
around the saturation intensity.

\section{Optical dipole guide}


\begin{figure}[htbp]
\begin{center}
\includegraphics[width=5cm]{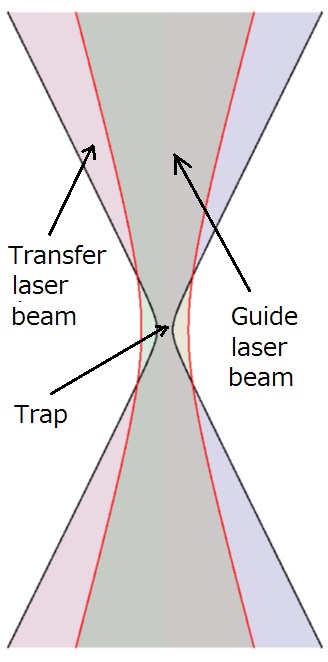}
\caption{Transverse cooling of the atomic beam.
The diameter of the transfer laser is approximately $30~\mu$m at the
focal point. The diameter of the guide laser can be equal or larger
than that of the transfer laser. When the diameter is larger, the neon beam
is guided inside the laser beam for a long distance. When small, 
the transverse velocity of the neon beam is adiabatically cooled, and
the neon atoms are released quickly from the guide laser potential.}
\label{focusedbeam}
\end{center}
\end{figure}

\begin{figure}[htbp]
\begin{center}
\includegraphics[width=12cm]{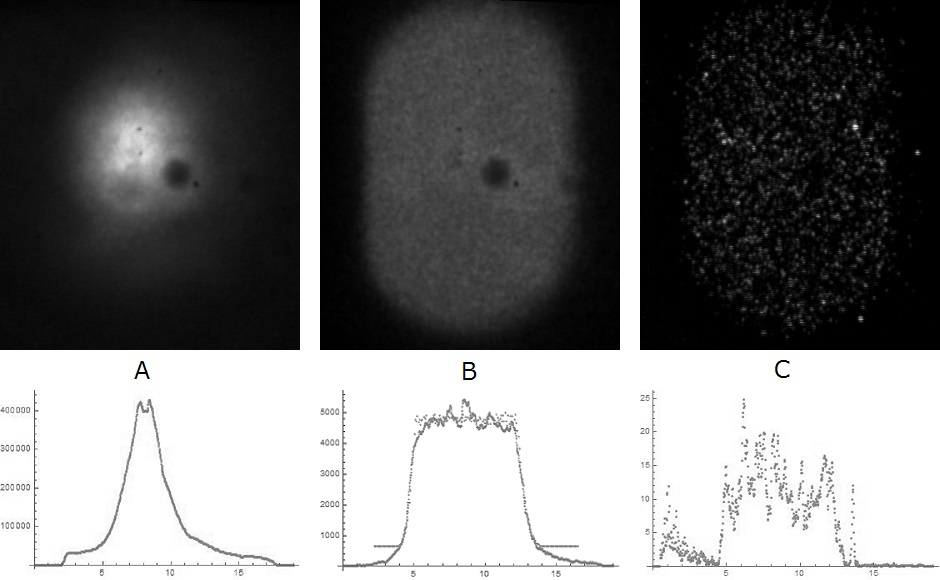}
\caption{Atomic beam pattern on the MCP. A: Pattern 
with the optical dipole guide.
The deflector laser is on to increase the metastable beam intensity.
B: optical dipole guide is turned off from A. C: both dipole guide and 
deflector
lasers are off. The brightness of each picture is
independently adjusted. The bottom curves show the intensity variation 
along the 
horizontal direction around the center. Ordinates show the real intensity 
ratio among the three pictures. Abscissa is in mm. 
Exposure time of pictures are 30~s. }
\label{beampattern}
\end{center}
\end{figure}


Since the transverse velocity is conserved, the atomic beam radius expands
as it falls down. The atomic beam can be compressed by guiding the beam
with a laser red-detuned to the  $1\mbox{s}_3-3\mbox{p}_5$
transition. We overlapped a red-detuned
laser on the transfer laser. The red-detune laser was focused to the same
spot as the focal point of the transfer laser (see Fig.~\ref{focusedbeam}). 
this red-detuned laser cools the transverse momentum of atoms in the beam,
and increases the atomic density in the beam up to several orders of magnitude.
Its function depends on the divergence and the intensity of the laser.
When the laser is tightly focused, falling atoms
feel radially expanding potential, and are cooled its radial motion.
When the radius at the focal point is sufficiently large, the laser will
 keep its radius for a long distance, and the atom is trapped
inside the laser beam.  

Figure~\ref{beampattern}a shows typical MCP images of the shadow of 
the gate for various operation conditions. 
C is the image obtained by a weak atomic source, B shows the case
of strong source, where the deflector-collimator is used. In Fig.~A optical
dipole laser focuses the 1s$_3$ atomic beam. The ordinate of the graphs
in the lower part shows the actual intensity ratio. From C to A, the atomic
beam intensity increases almost a factor of $10^4$.

\section{Example of diffraction pattern}


\begin{figure}[htbp]
\begin{center}
\includegraphics[width=8cm]{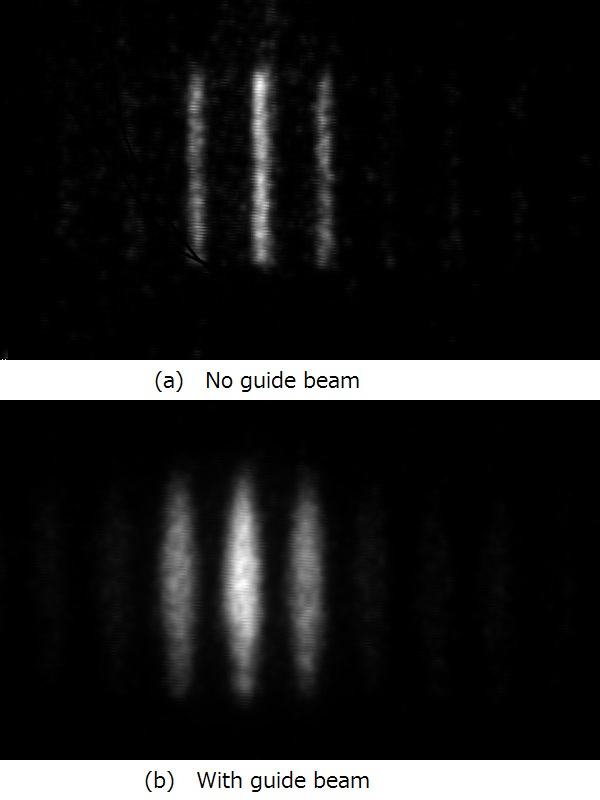}
\caption{The zero-th and $\pm 1$-st order diffraction patterns from a grating.
The grating size is $3 \times 0.2$~mm, Its pitch is 2~$\mu$m with 1~$\mu$m
slit. For a simple pattern such as this diffraction from a grating, the 
interferomrtric image 
can be recognized by real time video. When the atomic beam is guided
by laser, the image broadens, because the effective source size increases.
The exposure time is 5~s.}
\label{grating}
\end{center}
\end{figure}

\begin{figure}[htbp]
\begin{center}
\includegraphics[width=8cm]{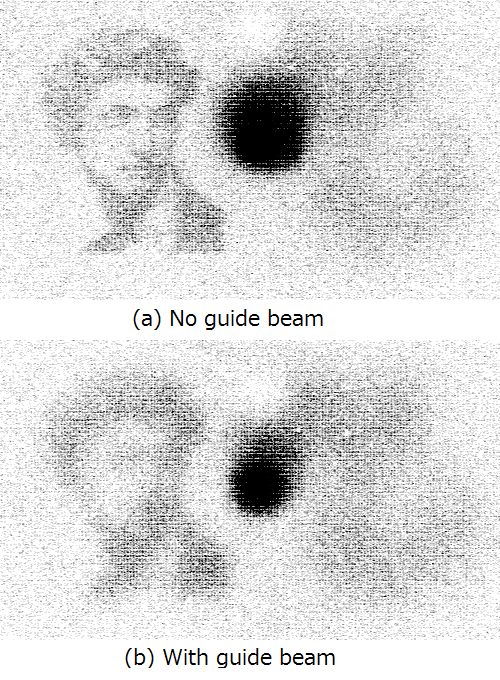}
\caption{The holographic pattern of reference \cite{juli}.
(a) is without dipole guide, and (b) is with dipole guide.
The exposure time is 120~min for (a), and 30~min for (b). 
The hologram size is 800~$\mu$m square with 
20~\% opening. To obtain diffraction pattern much more complicated than
Fig.~\ref{grating}, a long exposure time is necessary. Furthermore,
the image is degraded if the transverse cooling by an optical-dipolle guide
is used, due to the increased effective source size as shown in (b).
Meshes seen in the figure is artifacts of the video software(s).}
\label{juliano}
\end{center}
\end{figure}


We show in this section two interferometric patterns obtained with our atomic
beam described in the previous sections. 
Figure~\ref{grating} shows the zero-th and $\pm$ first order diffraction 
patterns from a narrow grating.
For a simple interferometric pattern like this case 
a single video frame of
1/30~s exposure time is sufficient to recognize gross pattern.
For a more complicated pattern such as the image from the computer generated
holographic image in  Fig~\ref{juliano} relatively long
accumulation time is necessary. Furthermore, degradation of the source
quality harms the quality of the picture as shown in Fig.~(b), which
uses the guide laser to increase the atomic beam intensity.

\section {Discussion}

Since metastable rare gas atoms can be detected with a high quantum efficiency,
they are ideal atomic species for the demonstration of interferometric 
phenomena. Unfortunately, metastable rare gas atoms have a large
inelastic collision rate. 
We demonstrated a bright neon atomic beam in the 1s$_3$ metastable state.
Atoms are released from the "magnetic
trap" which is formed near the center of quadrupole magnetic field of a MOT. 
At this moment we do not know the precise dynamics of the trapped atoms, 
such as, 
the range of atoms which is collected by the "magnetic trap", and how they are
cooled to the observed temperature. More precise analysis of dynamics
may bring to the generation of better cold atomic beams.

Another interesting experiment is to try with atoms which have smaller
inelastic collision rate. Application of known other cooling techniques
may bring a brighter cold atomic beam, or even finding of 
new collective phenomena.

\section*{Acknowledgements}

The author would like to thank Prof. J. Fujita for supplying us the holographic
film of Giuliano de Medici (Fig.~\ref{juliano}). Its pattern was calculated 
by Prof. T. Kishimoto.


\end{document}